%% file: doubleml_serverless_arxiv.tex
\documentclass[sigconf, nonacm]{acmart}

\ttfamily
\hyphenchar\font=`\-
\DeclareFontFamily{\encodingdefault}{\ttdefault}{\hyphenchar\font=`\-}

\usepackage{listings}
\usepackage{subfigure}

\begin{document}

\title{Distributed Double Machine Learning with a Serverless Architecture}

\author{Malte S. Kurz}
\affiliation{%
  \institution{University of Hamburg}
  \city{Hamburg}
  \country{Germany}
}
\email{malte.simon.kurz@uni-hamburg.de}

%\renewcommand{\shortauthors}{}

\begin{abstract}
This paper explores serverless cloud computing for double machine learning.
Being based on repeated cross-fitting, double machine learning is particularly well suited to exploit the high level of parallelism achievable with serverless computing.
It allows to get fast on-demand estimations without additional cloud maintenance effort.
We provide a prototype Python implementation \texttt{DoubleML-Serverless} for the estimation of double machine learning models with the serverless computing platform AWS Lambda and demonstrate its utility with a case study analyzing estimation times and costs.
\end{abstract}





\maketitle

\input{doubleml_serverless}

\begin{acks}
We are grateful for helpful comments by Philipp Bach, Martin Spindler and three anonymous referees.
This work was funded by the \grantsponsor{}{Deutsche Forschungsgemeinschaft (DFG, German Research Foundation)}{}
-- Project Number \grantnum{}{431701914}.
\end{acks}

\bibliographystyle{ACM-Reference-Format}
\bibliography{doubleml_serverless}

\end{document}

%% file: doubleml_serverless.tex
\section{Introduction}
Double machine learning (DML) models \cite{chernozhukov2018} are becoming increasingly popular among statisticians, econometricians and data scientists
with numerous methodological extensions \cite{Kallus2020,narita2020,klaassen2017,bach2020,colangelo2020,semenova2020,lewis2020,klaassen2018,chang2020}
and applications in areas like finance \cite{feng2020}, COVID-19 research \cite{chernozhukov2021,torrats2021} or economics \cite{knaus2020,semenova2020b}.
The DML models allow researchers to exploit the excellent prediction power of machine learning algorithms in a valid statistical framework for estimation and inference on causal parameters.
Recently, the Python and R packages \texttt{DoubleML} with a flexible object-oriented structure for estimating double machine learning models have been published \cite{DoubleML2020Py,DoubleML2020R}.

Serverless cloud computing is predicted to be the dominating and default architecture of cloud computing in the coming decade (Berkley View on Serverless Computing \cite{berkeley2019}) and is becoming increasingly adopted in the industry and by researchers.
Its Function as a Service (FaaS) paradigm lowers the entry bar to cloud computing technologies as the cloud providers are responsible for almost every operational and maintenance task.
A key advantage of serverless computing is the high elasticity in terms of an automated on-demand scaling depending on the actual amount of computing requests.
A second key advantage of serverless computing is the pricing model: Only actually used resources are charged without provisioning costs.

The management of computing clusters is usually not part of the daily business of econometricians or data scientists using DML for data analysis or in applied research.
Nevertheless there is demand for a high level of scalability to speed up the estimation of models like DML in interactive data analysis tasks.
%For example the comparison of different estimation methods, high-dimensional regression models, the estimation of heterogeneous treatment effects or high-dimensional instrumental variables are computationally intensive tasks when considering DML models.
In our experience, econometricians or data scientists who consider using cloud computing resources often want to achieve goals like the following:
\begin{itemize}
\item A high level of parallelism.
\item A ``cloud button'': Easy deployment and if possible no ongoing maintenance tasks for the user.
\item A high level of elasticity: On-demand availability of a high level of parallelism, pay-per-request and ideally no costs when the systems are idle.
\end{itemize}
The goal of this paper is to explore to what extent such goals are achievable with serverless cloud computing and we put special focus on DML models as an application.
Our study is based on AWS Lambda
and we made our prototype implementation \texttt{DoubleML-Serverless} publicly available.\footnote{GitHub: \url{https://github.com/DoubleML/doubleml-serverless} and AWS Serverless Application Repository: \url{https://serverlessrepo.aws.amazon.com/applications/eu-central-1/839779594349/doubleml-serverless}.}
We demonstrate the functionalities of the prototype with an experiment where we analyze estimation times and costs with different settings.

The rest of the paper is organized as follows:
Introductions to serverless computing and double machine learning are given in Sections \ref{secServerless} and \ref{secDML}.
The prototype implementation \texttt{DoubleML-Serverless} is described in Section \ref{secSDML}.
Section \ref{secExp} presents our experiment setup and results.
In Section \ref{secDisc} we discuss our prototype implementation and give an outlook to potential future extensions.
Section \ref{secConclusion} concludes the paper.

\section{Serverless Computing}\label{secServerless}
A core principle of serverless computing is that the user just writes a cloud function, often in a high-level programming language like Python, and all the server provisioning and administration is done by the cloud provider.
These serverless cloud function offerings are often called Function as a Service (FaaS), because the user basically only specifies the function code to be executed and declares which events should trigger such function calls.
There is especially no need for ex-ante provisioning of computing resources.
It is in the hand of the cloud provider to automatically scale up resources depending on the number of requests sent to the FaaS.
This is one of the key differences in comparison to a classical cloud server, where the user ex-ante needs to decide which requirements best match the upcoming computing tasks.

General discussions of serverless computing, recent developments and challenges can be found in \cite{berkeley2019,kuhlenkamp2020,baldini2017,hellerstein2018,vanEyk2017,vanEyk2018}.
Besides that, serverless computing is getting more and more adopted for various machine learning tasks, like for example to serve deep learning models  \cite{bhattacharjee2019,ishakian2018,tu2018} or more generally for ML model training and hyperparameter tuning \cite{carreira2018,wang2019,carreira2019}.

Another core principle of serverless computing is the pricing model.
The billing is usually done proportionally to the actually used resources and not proportionally to resources provisioned.
In case of AWS Lambda it is proportional to the execution time and very fine grained as the duration billing granularity was recently lowered to per millisecond billing \cite{aws1ms}.

When using AWS Lambda there is one key parameter set by the user, which is the memory available to the function at runtime.
AWS Lambda also scales other resources like CPU power proportionally to the allocated memory.
In the past the maximum memory allocatable was regularly increased and recently there was an significant extension from a maximum of $3$ GB to $10$ GB \cite{aws10gb}.
According to AWS this translates to a maximum of 6 vCPUs accessible in a single FaaS request \cite{aws10gb}.
By its nature the enormous elasticity of serverless computing platforms comes at the cost of rather strict resource limits for a single request.
When using AWS Lambda among others the maximum runtime is $15$ minutes.
However, the recent updates make serverless computing increasingly attractive for computationally intense tasks like machine learning.

\section{A Brief Introduction to Double Machine Learning}\label{secDML}
Double machine learning (DML) was developed in a series of papers \cite{belloni2014,belloni2015,belloni2018} and introduced as a general framework in \cite{chernozhukov2018}.
The application of DML for model classes like the partially linear regression model, the partially linear instrumental variable model, the interactive regression model and the interactive instrumental variable model is discussed in \cite{chernozhukov2018}.
Recently the DML framework and related techniques have been extended to numerous model classes like for example
reinforcement learning \cite{Kallus2020,narita2020},
transformation models \cite{klaassen2017},
generalized additive models \cite{bach2020},
continuous treatment effects \cite{colangelo2020,semenova2020},
dynamic treatment effects \cite{lewis2020},
Gaussian graphical models \cite{klaassen2018},
difference-in-differences models \cite{chang2020} and many more.
In these applications of DML, one is usually interested in statistical inference for a causal parameter $\theta_0$.
The DML framework makes it possible to obtain valid statistical inference for $\theta_0$ while exploiting the excellent prediction quality of machine learning methods for estimating nuisance functions denoted as $\eta_0$.

\begin{figure}[h]
\centering
\includegraphics[width=1.8in]{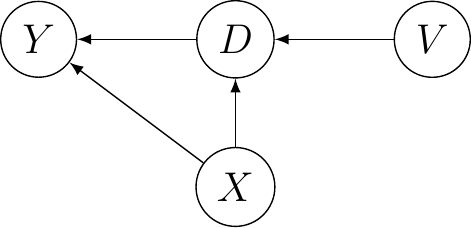}
\caption{Causal Diagram for the PLR Model \eqref{eq_plr_1}--\eqref{eq_plr_2}.}
\label{fig_causal_graph}
\end{figure}

As an example, we consider the partially linear regression (PLR) model as studied by \cite{robinson1988}
\begin{align}
Y &= D \theta_0 + g_0(X) + U, & \mathbb{E}(U|X, D) &= 0, \label{eq_plr_1}\\
D &= m_0(X) + V, & \mathbb{E}(V|X) &= 0, \label{eq_plr_2}
\end{align}
with outcome variable $Y$, treatment/policy variable $D$ and the potentially high-dimensional vector of controls $X := (X_1, \ldots, X_p)$.
The causal parameter of interest is $\theta_0$.
It measures the average treatment effect of $D$ on $Y$, if $D$ is conditionally exogenous.
The confounding variables $X$ affect $D$ via the function $m_0(X)$ and $Y$ via the function $g_0(X)$.
Figure \ref{fig_causal_graph} visualizes the interpretation in a causal diagram.
The DML framework allows to obtain valid statistical inference for $\theta_0$ while exploiting the excellent prediction quality of machine learning methods when estimating the nuisance functions $\eta_0=(g_0, m_0)$.
In the DML framework the nuisance functions $\eta_0=(g_0, m_0)$ can be estimated with different ML-methods, e.g., \cite{chernozhukov2018} use 
random forests, regression trees, boosting, lasso, neural networks and ensembles of these methods.
Depending on the structural assumptions on $\eta_0$, different ML-methods are appropriate.\footnote{We refer to \cite[Section 3]{chernozhukov2018} for a discussion and the formal conditions for the quality of the nuisance estimators.}

A key component of the DML framework are so-called Neyman orthogonal score functions $\psi(W; \theta, \eta)$.
The score functions identify the causal parameter of interest $\theta_0$ as the unique solution to $\mathbb{E}(\psi(W; \theta_0, \eta_0)) = 0$.
Neyman orthogonality of $\psi(W; \theta, \eta)$ with respect to the nuisance functions $\eta$ guarantees that there are no first-order effects of estimation errors in the nuisance functions on the estimation of the causal parameter $\theta_0$.

A second key component of the DML framework is sample splitting to avoid biases caused by overfitting.
The application of repeated cross-fitting is further recommended in \cite{chernozhukov2018}.
This makes it particularly well suited for a distributed architecture where the computationally intense inference tasks run in parallel.
Estimation of typical DML models often requires the estimation and prediction of several hundreds of ML models to approximate nuisance functions in different sample splits.
An ambitious goal of a serverless DML implementation would be to achieve that the estimation of the whole DML model with repeated cross-fitting does not take much longer than the estimation of the nuisance functions on a single fold.
The enormous elasticity of serverless cloud computing makes such a goal achievable in an on-demand setup with no need to start and maintain a large computing cluster, which is becoming costly if being idle.

The DML algorithm with repeated cross-fitting can be summarized as follows (w.l.o.g.\ we assume that the number of observations $N$ is divisible by the number of folds $K$): 
\begin{enumerate}
\item For each $m \in [M] := \lbrace 1, \ldots, M\rbrace$ draw a $K$-fold random partition $(I_{m,k})_{k \in K}$
of observation indices $[N] := \lbrace 1, \ldots, N\rbrace$ of size $n=N/K$.
Define $I_{m,k}^c := [N] \setminus I_{m,k}$ and for each $k$ construct a ML estimator
\begin{align*}
\hat{\eta}_{0,k} = \hat{\eta}_{0}((W_i)_{i \in I_{m,k}^c}).
\end{align*}
\item For each sample split, compute an estimate $\tilde{\theta}_{0,m}$ of the causal parameter as the solution to the equation
\begin{align*}
\frac{1}{N} \sum_{k=1}^{K} \sum_{i\in I_{m,k}} \psi(W_i; \tilde{\theta}_{0,m}, \hat{\eta}_{0,k}) = 0.
\end{align*}
The final estimate for the causal parameter is obtain via aggregation $\tilde{\theta}_{0} = \text{Median}((\tilde{\theta}_{0,m})_{m\in[M]})$.
\end{enumerate}
Note that the number of nuisance functions, which need to be estimated with ML methods, depends on the considered model, e.g., for the PLR model we have $L=2$ nuisance functions $\eta_0 = (g_0, m_0)$.
The total number of ML fits is $M \times K \times L$, i.e., one ML estimation in each fold, of each repeated sample splitting and for each nuisance function.
For example \cite{chernozhukov2018} choose $K=5$ (or $K=2$) and $M=100$, which for the PLR model with $L=2$ nuisance functions amounts to $1000$ (or $400$) ML fits or for the partially linear instrumental variable model with $L=3$ nuisance functions it amounts to $1500$ (or $600$) ML fits.
Note that for the interactive regression models, as considered in \cite{chernozhukov2018}, even more nuisance functions need to be estimated.
As mentioned before, our prototype for serverless DML allows for parallelization of all these $1000$ machine learning tasks and therefore potentially speeds up the estimation of DML models by a significant factor.
Basically, the estimation time with repeated cross-fitting with five folds and $100$ repetitions could be almost reduced to the time needed to estimate a single nuisance function for one fold in a single sample split.
Note that we do not require to transfer the estimated ML models for the nuisance functions $\hat{\eta}_{0,k}$, instead it suffices to return the predictions on the test datasets (i.e., for the observations indexed with $i \in I_{m,k}$) to evaluate the score function and solve for the causal parameter $\tilde{\theta}_{0,m}$ in a second step.

Neyman orthogonal score functions for many model classes, like for example the PLR model, can be written as linear functions in the parameter $\theta$, i.e.,
\begin{align*}
\psi(W; \theta; \eta) = \theta \psi_a(W; \eta)  + \psi_b(W; \eta).
\end{align*}
This common property forms the basis for a very general object-oriented implementation of DML models in the Python package \texttt{DoubleML} \cite{DoubleML2020Py}, which serves as a basis four our prototype \texttt{DoubleML-Serverless}.

\section{Serverless Double Machine Learning}\label{secSDML}
Similar to \texttt{PyWren} \cite{jonas2017}, our prototype implementation \texttt{DoubleML-Serverless} is intended to be used in an interactive fashion:
The user runs a Python session on a local machine or server, but at the same time has access to a high level of parallelism with an on-demand and pay-per-request interface for the computationally most intense tasks during the estimation of DML models.\footnote{
As suggested by an anonymous referee, alternatively a fully serverless version could be implemented using services like AWS Step Functions to organize the serverless workflow.}
In comparison to \texttt{PyWren}, which allows to run more or less arbitrary parallel tasks, like for example  map reduce, our implementation is more specialized to the specific use case of DML models.
Many cloud providers have serverless FaaS offerings.
Our prototype \texttt{DoubleML-Serverless} uses AWS Lambda and is developed in Python as an extension of the \texttt{DoubleML} package \cite{DoubleML2020Py}.\footnote{
The prototype is tied to AWS Lambda. Adaptions of the data transfer and the deployment process would be necessary to make it compatible with other serverless platforms.}

\subsection{The Architecture of DoubleML-Serverless}
\begin{figure}[h]
\centering
\includegraphics[width=3.3in]{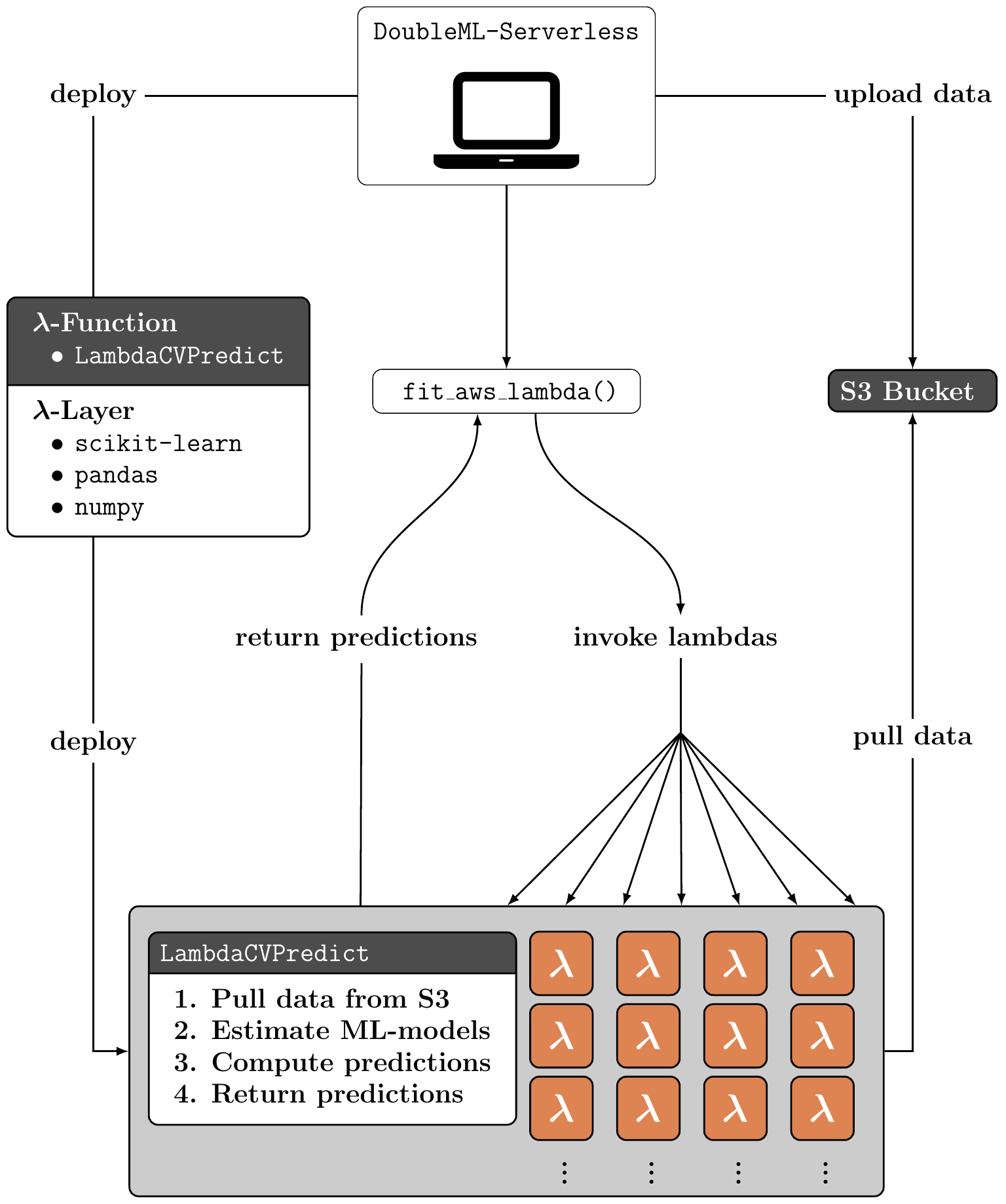}
%\caption{\texttt{DoubleMLPLRServerless}: Architecture and the level of scaling.}
\caption{DoubleML-Serverless: Architecture.}
\label{fig_serverless_architecture}
\end{figure}
The architecture of \texttt{DoubleML-Serverless} is summarized in Figure \ref{fig_serverless_architecture}.
As data storage we use the AWS S3 object storage.
In the \texttt{DoubleML-Serverless} package, we implement a \texttt{DoubleMLDataS3} class, which serves as a data backend.
It is inherited from the \texttt{DoubleML} class \texttt{DoubleMLData} and primarily extends it by methods to transfer datasets from and to AWS S3.
The model classes, like for example \texttt{DoubleMLPLRServerless} for the PLR model, extend the corresponding classes
%(\texttt{DoubleMLPLR} for the partially linear regression model)
from the \texttt{DoubleML} package by methods to perform the ML estimation and prediction step on AWS Lambda.
In addition to the standard inputs for \texttt{DoubleML} model classes, the user needs to provide the name of the deployed lambda function and the AWS region on initialization.
Then the DML model can be estimated with a call to the method \texttt{fit\_aws\_lambda()}.
On invocation, each request consists of a reference to the dataset on S3, the nuisance-function-specific names of target variables and confounders and the sample splitting.
The lambda function returns the predictions for the corresponding test indices.

\subsection{The Level of Scaling}
\begin{figure}[h]
\centering
\includegraphics[width=3.3in]{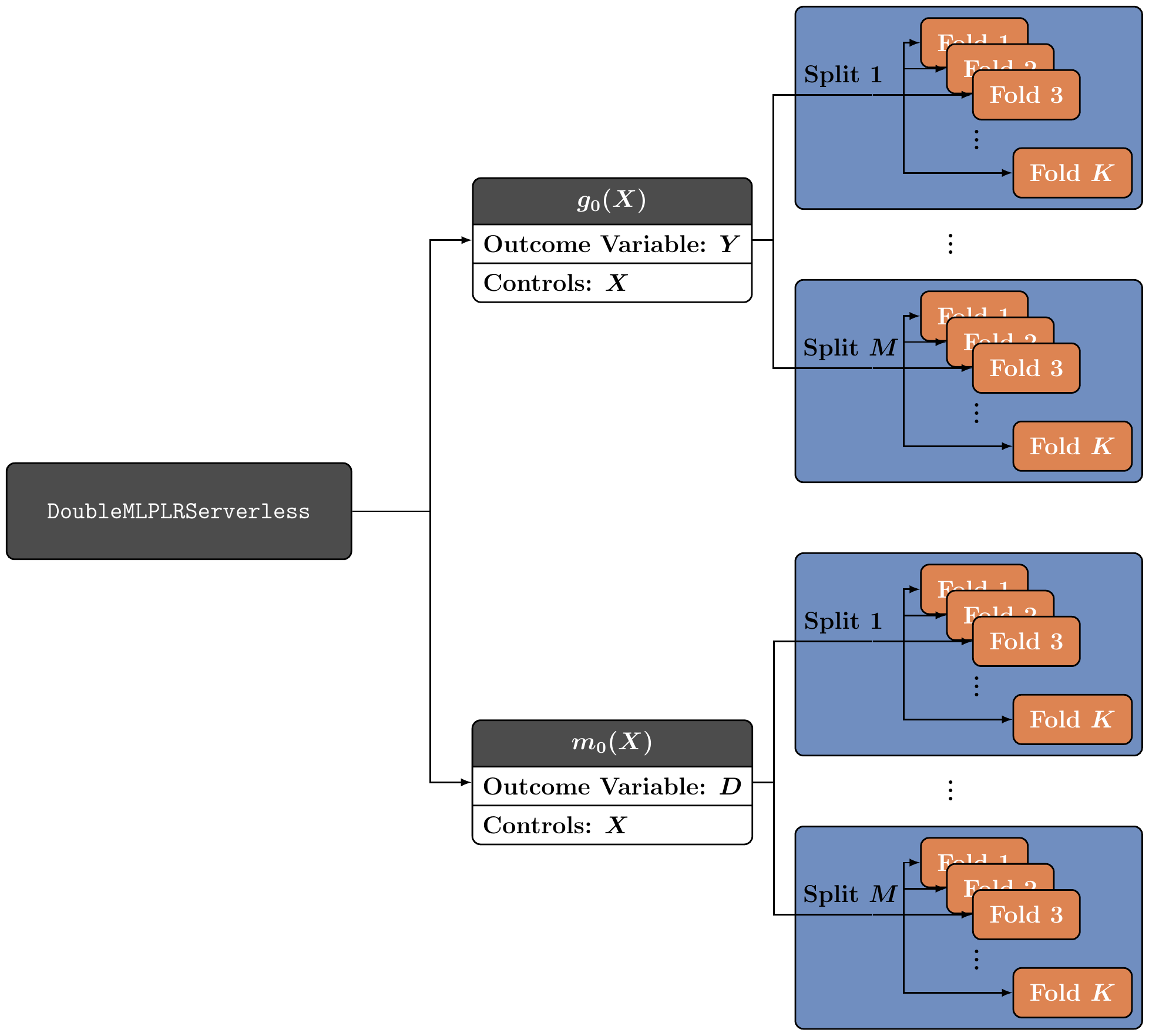}
%\caption{\texttt{DoubleMLPLRServerless}: Architecture and the level of scaling.}
\caption{DoubleML-Serverless: Level of Scaling.}
\label{fig_serverless_scaling}
\end{figure}

Our prototype implementation \texttt{DoubleML-Serverless} offers two different degrees of scaling.
Figure \ref{fig_serverless_scaling} visualizes the level of scaling options for the PLR class \texttt{DoubleMLPLRServerless}.
Per-sample-split scaling is achieved by choosing \texttt{scaling = 'n\_rep'}.
It results in a lambda function invocation for each nuisance function and repeated sample split, i.e., for each blue rectangle in Figure \ref{fig_serverless_scaling}.
In each such invocation, $K$ machine learning models are estimated and corresponding predictions for the test indices returned.
As an alternative one can choose \texttt{scaling = 'n\_folds * n\_rep'} to invoke a separate lambda for each single fold, nuisance function and sample split, i.e., for each orange rectangle in Figure \ref{fig_serverless_scaling}.

If we again consider the above mentioned PLR model with $K=5$ folds, $M=100$ splits and $L=2$ nuisance functions, it means that we either sent $M \times L = 200$ requests or $M \times K \times L = 1000$ requests.
Which level of scaling is favorable depends on the individual use case.
First of all, the runtime limit of AWS Lambda implies that the per-sample-split scaling cannot be applied if the estimation of $K$ machine learning models takes longer than the maximum runtime, which might be the case, depending on the machine learning approach and the size of the dataset.
Furthermore, there is always a cost vs.\ estimation-time tradeoff which the user controls via the \texttt{scaling} parameter and the allocated memory.

\subsection{Deployment with AWS SAM}
User-friendly deployment of the prototype is achieved with the AWS Serverless Application Model (AWS SAM) for deploying our FaaS to AWS Lambda.
AWS SAM \cite{awsSam} allows for easy deployment of serverless applications to AWS Lambda and is configured via template files.
We added an AWS SAM template to our prototype, which deploys the following components (see Figure \ref{fig_serverless_architecture} for a visualization of the architecture):
\begin{itemize}
\item A lambda function called \texttt{LambdaCVPredict}.
\item A layer providing the Python libraries \texttt{scikit-learn}, \texttt{pandas} and \texttt{numpy} together with their dependencies.
\item An S3 bucket for the data transfer (can be optionally generated, or an existing bucket is used).
\item A role for the execution of the lambda function \texttt{LambdaCVPredict} which consists of the AWS-managed \texttt{AWSLambdaBasicExecutionRole} policy plus read access to the S3 bucket for data transfer.
\end{itemize}
\texttt{LambdaCVPredict} is the main function being invoked when estimating DML models on AWS Lambda.
The main advantage of AWS SAM is that the deployment process is simple with only two calls \texttt{sam build} and \texttt{sam deploy --guided}.
Additionally, based on the same SAM template, even simpler deployment is offered directly from the AWS Serverless Application Repository.\footnote{\url{https://serverlessrepo.aws.amazon.com/applications/eu-central-1/839779594349/doubleml-serverless}}
The listing in the AWS Serverless Application Repository gives the user almost a ``bring me to the cloud''-button for estimating DML models.\footnote{From the AWS Serverless Application Repository, the deployment can be done directly in the browser by clicking ``Deploy'' and following the steps in the AWS Management Console.}

\section{Estimating Double Machine Learning Models with DoubleML-Serverless}\label{secExp}
To demonstrate our prototype implementation \texttt{DoubleML-Serverless} we revisit the Pennsylvania Reemployment Bonus experiment and estimate the effect of provisioning a cash bonus on the unemployment duration as studied in \cite{chernozhukov2018}.

\subsection{Experiment Setup}
We consider the previously discussed PLR model \eqref{eq_plr_1}--\eqref{eq_plr_2}.
The nuisance functions $g_0$ and $m_0$ are estimated using a random forest with $500$ regression trees.\footnote{
In case of a binary treatment variable $D$, one can also use classifiers to estimate $m_0$.}
We choose $K=5$ folds and $M=100$ splits.

At invocation, the following information is transferred to \texttt{LambdaCVPredict}:
\begin{itemize}
\item The name of the outcome variable, e.g., for $g_0$ the $Y$ column.
\item The names of the controls, e.g., for $g_0$ the $X$ columns.
\item The ML model to be estimated, e.g., random forest.
\item The set of indices $I_{m,k}$.
\end{itemize}
In Listing \ref{sample_code} we provide sample code which demonstrates the syntax to estimate the described DML model with \texttt{DoubleML-Serverless}
for the \textit{bonus} dataset.
\begin{lstlisting}[caption=Estimation of a Partially Linear Regression (PLR) Model with DoubleML-Serverless., label=sample_code]
from doubleml.datasets import fetch_bonus
from doubleml_serverless import DoubleMLDataS3, DoubleMLPLRServerless
from sklearn.base import clone
from sklearn.ensemble import RandomForestRegressor

df_bonus = fetch_bonus('DataFrame')
dml_data_bonus = DoubleMLDataS3(
    'doubleml-serverless-data', 'bonus_data.csv', df_bonus,
    y_col='inuidur1', d_cols='tg',
    x_cols=['female', 'black', 'othrace', 'dep1', 'dep2',
       'q2', 'q3', 'q4', 'q5', 'q6', 'agelt35',
       'agegt54', 'durable', 'lusd', 'husd'])
dml_data_bonus.store_and_upload_to_s3()

ml = RandomForestRegressor(n_estimators = 500, n_jobs=-1)
ml_g = clone(ml)
ml_m = clone(ml)
dml_lambda_plr_bonus = DoubleMLPLRServerless(
    'LambdaCVPredict', 'eu-central-1',
    dml_data_bonus, ml_g, ml_m,
    n_folds=5, n_rep=100)

dml_lambda_plr_bonus.fit_aws_lambda()
\end{lstlisting}

%The outcome variable, controls and the ML model depend on the nuisance function which should be estimated.
The FaaS function \texttt{LambdaCVPredict} returns predictions which are obtained by estimating the nuisance function based on the training indices $I_{m,k}^{c}$ and then predictions are computed for all $i \in I_{m,k}$.
When all requested predictions have been returned, the score function components for the PLR model at hand are obtained as
\begin{align*}
\psi_a(W_i; \hat{\eta}_0) &= - (D_i - \hat{m}_0(X_i)) (D_i - \hat{m}_0(X_i)), \\
\psi_b(W_i; \hat{\eta}_0) &= (Y_i - \hat{g}_0(X_i))(D_i - \hat{m}_0(X_i)).
\end{align*}
Using the evaluated score function components, we can solve for the parameter estimate
\begin{align*}
\tilde{\theta}_0 = \frac{- \sum_{i=1}^{N} \psi_b(W_i; \hat{\eta}_0)}{\sum_{i=1}^{N} \psi_a(W_i; \hat{\eta}_0)}.
\end{align*}
Based on the evaluated score function, inference tasks like the computation of standard errors and confidence intervals that build on a multiplier bootstrap approach could be easily done locally using the functionalities of the \texttt{DoubleML} package.
For further details, we refer to the paper introducing the DML framework \cite{chernozhukov2018} and the documentation of the \texttt{DoubleML} package \cite{DoubleML2020Py}.\footnote{\url{https://docs.doubleml.org}}

\subsection{Timings and Costs}
\begin{figure}[h]
\centering
\includegraphics[width=3.2in]{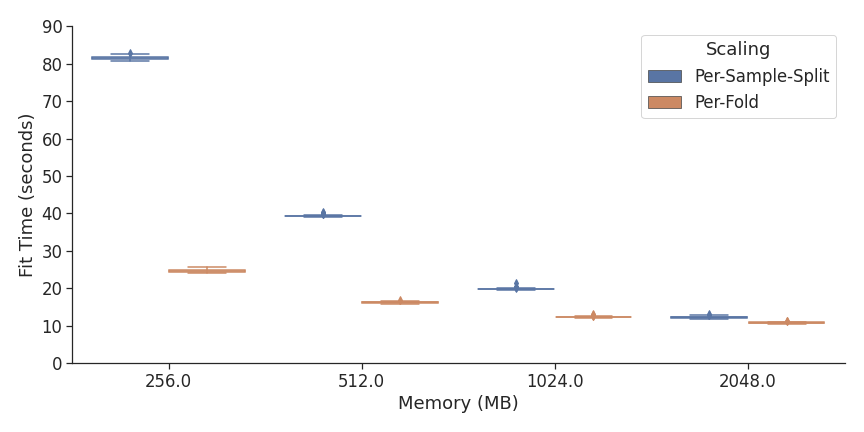}
\caption{Serverless Fit Times with Different Scaling and Allocated Memory.}
\label{fit_time}
\end{figure}

\begin{figure}[h]
\centering
\includegraphics[width=3.2in]{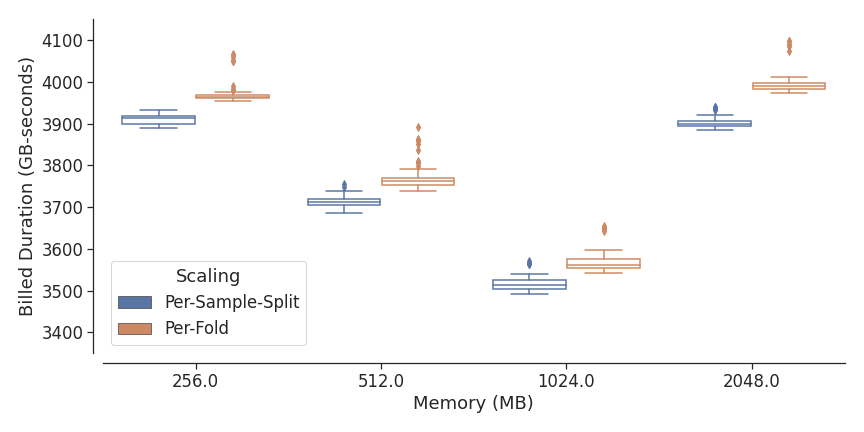}
\caption{Serverless Costs with Different Scaling and Allocated Memory.}
\label{costs}
\end{figure}

%\begin{figure*}
%\subfigure[Fit Time in seconds]{\includegraphics[width=3.2in]{fit_time}
%\label{fit_time}}
%\subfigure[Costs in GB-seconds]{\includegraphics[width=3.2in]{costs}
%\label{costs}}
%\caption{Serverless Fit Times and Costs with Different Scaling and Allocated Memory.}
%\label{fig_sim}
%\end{figure*}
To demonstrate the utility of our prototype \texttt{DoubleML-Serverless} we ran a couple of experiments on AWS Lambda with the above stated \textit{bonus} data example.
We especially focus on the two different settings for the \texttt{scaling} parameter, i.e.,
\texttt{scaling = 'n\_rep'} for per-sample-split scaling and \texttt{scaling = 'n\_folds * n\_rep'} for per-fold scaling.
With the above mentioned settings ($K=5$ folds and $M=100$ splits) this amounts to $200$ and $1000$ invocations, respectively.
Additionally, we also alter the memory available to the function at runtime which also impacts the CPU power, because AWS Lambda scales other resources proportionally to the allocated memory.
All experiments are repeated $100$ times and the estimation times and costs are visualized with boxplots in Figure \ref{fit_time} and \ref{costs}.

In Figure \ref{fit_time} we can clearly see that the total estimation times for the DML models decrease if more memory is allocated.
However, the marginal improvement in the estimation times is decreasing which is a typical behavior as for example documented in \cite{akhtar2020,ishakian2018}.
It is also important to point out that faster estimation does not necessarily come at higher costs.
In Figure \ref{costs} we see that by allocating more memory, $512$ MB or $1024$ MB instead of $256$ MB, besides lowering the estimation time we could also lower the total costs for the estimation on AWS Lambda.
The observation that too low or high memory allocations result in higher costs is also common for serverless computing with AWS Lambda and this observation has been used to propose cost optimization frameworks  \cite{akhtar2020,casalboni}.

When comparing the two different levels of scaling, we can see in Figure \ref{fit_time} that by choosing per-fold scaling the estimation times can be further decreased.
It is important to note that the costs are only slightly increasing when going from per-sample-split to per-fold scaling (see Figure \ref{costs}).
This is one of the benefits of serverless computing where one can increase the concurrency dramatically but still the billing is proportional to the actual computing time and therefore is often only slightly increased due to more overhead.

Table \ref{table_timing_example} provides more detailed results for the cheapest case in our experiment which is the setting with $1024$ MB memory allocated and per-sample-split scaling.
We can see that in the $100$ repetitions of our experiment the estimation time was on average $19.82$ seconds.
The response time from the invocation of the first lambda until we received the predictions from each of the $200$ invocations took on average $19.09$ seconds and the average computation time for a single invocation was $17.16$ seconds.\footnote{
The maximum total response time of 20.76 seconds for 200 invocations, each with an computation time between $17.16$ and $17.44$ seconds (see Table \ref{table_timing_example}), also gives some indication that a high level of elasticity seems to be achievable.
For an empirical evaluation of the elasticity of different FaaS platforms we refer to \cite{kuhlenkamp2020b}.}
Therefore, in this setting we are very close to the ambitious goal that using serverless computing the estimation of the DML model with repeated cross-fitting only takes a little bit more time than estimating with only a single sample split on a machine with similar CPU power as one lambda.
In Table \ref{table_timing_example} we can further see that the average estimation costs amount to $3515.36$ GB-seconds, which translates to roughly $0.05858$ USD at the current price of $0.0000166667$ USD per GB-second that AWS charges in eu-central-1 \cite{aws-lambda-pricing}.\footnote{
For comparison, the estimation of the same DML model on a virtual machine (AWS EC2 instance of type \texttt{m5.2xlarge} with 8 vCPUs) takes much longer with approximately $383.90$ seconds and at the same time amounts to slightly lower costs of $0.04905$ USD at the current price of $0.46$ USD per hour when ignoring the additional costs from setup and teardown of the virtual machine.}

\begin{table}[h]
\renewcommand{\arraystretch}{1.3}
\caption{Serverless Fit Times and Costs with 1024 MB Memory and Per-Sample-Split Scaling
(Mean, Min \& Max in 100 Runs).}
\label{table_timing_example}
\centering
\begin{tabular}{l || c | c | c}
\hline
& \bfseries Mean & \bfseries Min & \bfseries Max \\
\hline\hline
\bfseries Fit Time (s) & 19.82 & 19.53 & 21.49 \\
\bfseries Billed Duration (GB-s) & 3515.36 & 3492.01 & 3571.42 \\
\bfseries Avg.\ Duration per Invocation (s) & 17.16 & 17.05 & 17.44 \\
\bfseries Total Response Time (s) & 19.09 & 18.81 & 20.76 \\
\hline
\end{tabular}
\end{table}

\section{Discussions}\label{secDisc}
In the following, we discuss features, advantages and limits of the current prototype implementation \texttt{DoubleML-Serverless} and give an outlook to potential future extensions.

\textbf{Reproducibility and seeds:}
The prototype comes with a basic implementation of seeds to obtain reproducible results.
We refer to the \texttt{numpy} documentation \cite{numpy-rng} for a discussion of parallel random number generation.

\textbf{Launch overhead \& cold vs. warm invocations:}
It is well known that there is a launch overhead when using serverless computing which results in timing differences between so-called cold and warm starts.
We report timings for warm starts and refer to \cite{ishakian2018,berkeley2019} for a discussion of the phenomenon.

\textbf{Transfer of ML models:}
The ML models are transferred at invocation using their string representation and only a subset of all \texttt{scikit-learn} ML-models is supported.
To transfer more sophisticated learners, an alternative approach like pickling the learners similar to \texttt{PyWren} could be implemented.

\textbf{Data transfer via payloads:}
The prototype uses the payloads to transfer the test indices and to return the predictions.
This implies some restrictions, which could be overcome by implementation of a data transfer via S3.

\textbf{Distributed storage:}
The datasets, which are loaded in every learning task, are stored in the Amazon S3 object storage.
An alternative would be the AWS Elastic File System (EFS) which can be mounted directly for AWS Lambda calls \cite{aws-efs}.

\textbf{Cost optimization:}
The main configuration parameter of AWS Lambda is the allocated memory.
It is important to know that AWS Lambda allocates CPU power proportional to the amount of memory.
Therefore, the memory allocation has an impact on the total execution time and the costs.
Discussions and proposal for cost optimization of serverless applications are provided in \cite{ishakian2018,wang2019} and implementations of frameworks for cost optimization in \cite{akhtar2020,casalboni}.
Similar approaches could also be used to cost-optimize our prototype \texttt{DoubleML-Serverless}.

\textbf{Limits on runtime and memory:}
Currently on AWS Lambda, there is an upper limit for execution time of $15$ minutes.
Obviously, our prototype cannot be used if the single fold estimation is not doable within this limit.
Considering the previously discussed scenario with $K=5$ folds, $100$ splits and two nuisance functions and assuming that the estimation of each task is of similar effort, this translates to a total estimation time limit of roughly $10.5$ days ($5 \times 100 \times 2 \times 15$ minutes).
Note that in the past AWS Lambda regularly increased these limits.

\textbf{Limits on memory:}
Recently AWS Lambda announced a significant increase of their memory limit from $3$ GB to $10$ GB \cite{aws10gb}.
This implies that serverless computing is becoming increasingly suitable and attractive for memory-intense models and big data applications.
For standard applications of DML these memory limits are not an issue.
However, DML is particularly well suited for causal inference in high-dimensional settings and therefore also used for very big datasets.
Realizing such estimations in very high-dimensional and big data sets with our prototype will be challenging.

\textbf{Parameter tuning for DML models:}
As usual in machine learning, hyperparameter tuning is also done for DML models.
The prototype could be extended to also support hyperparameter tuning with an efficient serverless implementation.

\textbf{DML models with multiple treatment variables:}
The prototype implementation only supports a single treatment variable but an extension to multiple treatment variables, as supported by \texttt{DoubleML}, would be straightforward.

\section{Conclusion}\label{secConclusion}
For many users like econometricians, statisticians and data scientists existing serverfull frameworks for distributed machine learning have a high entry barrier and are often expensive if being used infrequently or inefficiently.
In this paper we explore serverless cloud computing for estimation of double machine learning models.
Our prototype \texttt{DoubleML-Serverless} using AWS Lambda gives econometricians, statisticians and data scientists access to an enormous level of parallelism, it almost comes with a ``cloud button'' as it can be easily deployed via AWS SAM and it comes at the advantage of a pay-per-request pricing model.